%% file: paper.tex
\documentclass[runningheads,a4paper]{llncs}

\usepackage{microtype}
\usepackage{url}                  
\usepackage{listings}          
\usepackage{enumitem}      

\usepackage{graphicx}
\usepackage{float,subfig}
\usepackage{xspace,framed}
\usepackage{colortbl}
\usepackage{calc}
 
\usepackage{algorithm}
\usepackage{amsfonts}
\usepackage{amsmath}
\usepackage{multicol}
\usepackage{multirow}
\usepackage{tikz}
\usepackage[justification=centering]{caption}
\usepackage{stmaryrd}
\usetikzlibrary{positioning, arrows, automata, shapes}
\usepackage{hhline}
\usepackage{pifont}
\usepackage{cite}
\usepackage{pdflscape} 
\usepackage{longtable}
\usepackage{afterpage}
\usepackage{wasysym}

\usepackage[justification=centering]{caption}
\input{macros}

\allowdisplaybreaks

\newcommand*{\extended}{}

\begin{document}





\title{\tool: Safe Semantic Refactoring to Java Streams}

\author{
\mbox{Cristina David}\inst{1} \and
Pascal Kesseli\inst{2} \and
Daniel Kroening\inst{2}}

\institute{University of Cambridge, UK 
\and
  University of Oxford, UK}




\maketitle

\begin{abstract}
  Refactorings are structured changes to existing software that leave its
  externally observable behaviour unchanged.  Their intent is to improve
  readability, performance or other non-behavioural properties. 
  State-of-the-art automatic refactoring tools are {\em syntax}-driven and,
  therefore, overly conservative.  In this paper we explore {\em
  semantics}-driven refactoring, which enables much more sophisticated
  refactoring schemata.  As~an exemplar of this broader idea, we present
  \tool, an automatic refactoring tool that transforms Java with
  external iteration over collections into code that uses Streams, a new
  abstraction introduced by Java~8.  Our refactoring procedure performs
  semantic reasoning and search in the space of possible refactorings using
  automated program synthesis.  Our experimental results support the
  conjecture that semantics-driven refactorings are more precise and are
  able to rewrite more complex code scenarios when compared to syntax-driven
  refactorings.
\end{abstract}





\section{Introduction}

Refactorings are structured changes to existing software which leave its
externally observable behaviour unchanged.  They improve non-functional
properties of the program code, such as testability, maintainability and
extensibility while retaining the semantics of the program.  Ultimately,
refactorings can improve the design of code, help finding bugs as well
as increase development speed and are therefore seen as an integral part
of agile software engineering processes~\cite{DBLP:conf/xpu/Kerievsky04b,
Fowler1999}.

However, manual refactorings are a costly, time-intensive, and not least
error-prone process.  This has motivated work on automating specific
refactorings, which promises safe application to large code bases at low
cost.  We differentiate in this context between {\em syntax-driven} and {\em
semantics-driven} refactorings.  While the former address structural changes
to the program requiring only limited information about a program's
semantics, the latter require detailed understanding of the program
semantics in order to be applied soundly.  An example of a refactoring that
requires a semantics-driven approach is {\em Substitute Algorithm}, where an
algorithm is replaced by a clearer, but equivalent
version~\cite{Fowler1999}.
%
%
A syntax-driven approach is insufficient to perform such substantial
transformations.  Figure~\ref{ex:syntax-limits} illustrates this using an
example: Both loops in the code implement the same behaviour.  In order to
recognise this and apply {\em Substitute Algorithm}, pattern-based
approaches need explicit patterns for vastly different syntaxes implementing
the same semantics, which is infeasible for practical applications.


Notably, the limitations of syntax-driven refactorings have been observed
in several works, resulting in an emerging trend to incorporate more {\em semantic}
information into refactoring decisions, such as Abstract Syntax Tree (AST) 
type information,
further preventing compilation errors and behaviour changes
\cite{Steimann2011,Steimann2012Pilgrim,Steimann2011KollePilgrim}.

\begin{figure}
  \begin{lstlisting}[mathescape=true,showstringspaces=false]
List<Integer> org = getData();
List<Integer> copy = new ArrayList<>();
for (int i=0; i < org.size(); ++i)
  if (org.get(i) > 0) copy.add(2 * org.get(i));

Iterator<Integer> it=org.iterator();
while (it.hasNext()) {
  int tmp = it.next() * 2;
  if (tmp <= 0) continue;
  copy.add(tmp);}
\end{lstlisting}
\caption{Limitations of pattern-based refactorings.}
\label{ex:syntax-limits}
\end{figure}

In this paper, we take a step further in this direction by proposing a
fully semantic refactoring approach.  There is a very broad space of
methods that are able to reason about program semantics.  The desire
to perform refactorings safely suggests the use of techniques that
overapproximate program behaviours.  As~one possible embodiment of
semantics-driven refactoring, we leverage software verification
technologies with the goal of reliably automating refactoring
decisions based on program semantics, as in the case of the {\em
  Substitute Algorithm} refactoring. Our research hypothesis is that
semantics-driven refactorings are more precise and can handle more
complex code scenarios in comparison with syntax-driven refactorings.

\paragraph{Demonstrator: Refactoring Iteration over Collections}

We use a particular refactoring as demonstrator for our idea. Nearly every
modern Java application constructs and processes collections.  A~key
algorithmic pattern when using collections is iteration over the contents of
the collection.  We~distinguish {\em external} from {\em
internal} iteration.

To enable external iteration, a Collection provides the means to enumerate
its elements by implementing Iterable.
Clients that use an external iterator must advance the traversal and
request the next element explicitly from the iterator.  External
iteration has a few shortcomings:
\begin{itemize}
\item Is inherently sequential, and must process the elements in the order
specified by the collection. This bars the code from using concurrency
to increase performance.
\item Does not describe the intended functionality, only that each element
is visited. Readers must deduce the actual semantics, such as finding an element
or transforming each item, from the loop body.
\end{itemize}


The alternative to external iteration is internal iteration, where
instead of controlling the iteration, the client passes an operation to
perform to an internal iteration procedure, which applies that
operation to the elements in the collection based on the algorithm it
implements. Examples of internal iteration patterns include finding an
element by a user-provided predicate or transforming each element
in a list using a provided transformer.
In order to enable internal iteration, Java~SE~8 introduces a new
abstraction called {\em Stream} that lets users process data in a
declarative way.  The {\em Stream} package provides implementations of
common internal iteration algorithms such as {\em foreach}, {\em find}
and {\em sort} using optimised iteration orders and even concurrency
where applicable.  Users can thus leverage multicore architectures
transparently without having to write multithreaded code.  Internal
iterations using {\em Stream} also explicitly declare the intended
functionality through domain-specific algorithms.  A call to Java~8
{\em find} using a predicate immediately conveys the code's intent,
whereas an externally iterating \code{for} loop implementing the same
semantics is more difficult to understand.
Figures~\ref{ex:find-query} (a) and (b) illustrate this
difference for the same {\em find} semantics.  Finally, external
iteration using a \code{for} loop violates Thomas' {\em DRY} principle
({\em ``Don't repeat yourself''} \cite{thomas}) if the intended
functionality is available as a {\em Stream} template.  Internal
iteration through {\em Stream} thus eliminates code duplication.

\begin{figure}
  \begin{lstlisting}[mathescape=true]
  Integer result = null;
  List<Integer> data = getData();
  for (int el : data)
    if (el % 2 == 0) {
      result = el; 
      break;}
\end{lstlisting} §\hspace{24em}\small{(a)}§
  \begin{lstlisting}[mathescape=true]
  List<Integer> newList = getData();
  Optional<Integer> result = list.stream()
                          .filter(el -> el % 2)
                          .findFirst();
\end{lstlisting} §\hspace{24em}\small{(b)}§
\caption{Find element in list with external (a) vs. internal (b) iteration.}
\label{ex:find-query}
\end{figure}

For illustration, consider the example in Fig.~\ref{ex:stream}~(a). 
This example uses external iteration to create a new list by multiplying
all the positive values in the list \code{list} by 2.  In this variant of the
code, we use a \code{while} loop to sequentially process the elements in the
list.

In Fig.~\ref{ex:stream} (b), we have re-written the code using streams. 
This variant of the code does not use a loop statement to iterate through the
list.  Instead, the iteration is done internally by the stream.  Essentially, we
create a stream of Integer objects via \code{Collection.stream()}, filter it
to produce a stream containing only positive values, and then transform it
into a stream representing the doubled values of the filtered list.

\begin{figure*}
\begin{minipage}{\textwidth}
  \begin{lstlisting}[mathescape=true,escapechar={§}]
  Iterator<Integer> it=list.iterator();
  List<Integer> newList=new ArrayList<Integer>();
  while (it.hasNext()) {
    int el=it.next().intValue();
    if (el > 0)
	newList.add(2 * el);
  } §\hspace{24em}\small{(a)}§
\end{lstlisting}
\end{minipage}\\
\begin{minipage}{\textwidth}
  \begin{lstlisting}[mathescape=true,escapechar={§}]
  List<Integer> newList=new ArrayList<Integer>();
  newList=list.stream()
             .filter(el -> el>0)
             .map(el -> new Integer(2 * el));
             .collect(toList());
  return newList; §\hspace{16em}\small{(b)}§
\end{lstlisting}
\end{minipage}
\caption{Filtering and mapping example with external (a) vs. internal (b) iteration.}
\label{ex:stream}
\end{figure*}

\paragraph{Goal of the paper} 
In this paper, we are interested in refactoring Java code handing
collections through external iteration to use streams. Our refactoring
procedure is based on the program semantics and makes use of program
synthesis.

\paragraph{Contributions:}
\begin{itemize}
\item We present a program synthesis based refactoring procedure for Java
code that handles collections through external loop iteration.
%
%
\item We have implemented our refactoring method in the tool \tool. Our
experimental results support our conjecture that semantics-driven
refactorings are more precise and can handle more complex code scenarios
than syntax-driven refactorings.
\end{itemize}



\section{Preliminaries} \label{sec:preliminaries}


%

\paragraph{General refactorings}
As we want to preserve generality, 
we are interested in  refactorings that are
correct independent of their context.  To motivate our decision, let's
look at the example in Fig.~\ref{ex:filter1}.  We define a method
\code{removeNeg} that removes the negative values in the list received as
argument, which we later call for the list \code{data}.  However, given that
\code{data} contains only positive values, applying \code{removeNeg} does not
have any effect.

\begin{figure}
  \begin{lstlisting}[mathescape=true,showstringspaces=false]
void removeNeg(ArrayList<Integer> l) {
  Iterator<Integer> it = l.iterator();
  while (it.hasNext())
    if (it.next() < 0) it.remove();
}
List<Integer> data = new ArrayList<>();
Collections.addAll(data, 1, 2, 3);
removeNeg(data);
\end{lstlisting}
\caption{Filter example.}
\label{ex:filter1}
\end{figure}

Thus, for this particular calling context, 
we could refactor the body of \code{removeNeg} to a NO-OP.
While this refactoring is correct for the code given in Fig.~\ref{ex:filter1}, 
it may cause problems during future evolution of the code 
as someone might use it for its original intended functionality 
(that of removing negative values). As we envision that our refactoring procedure 
will be used during the development process, we choose to not perform such 
strict refactorings.

\section{Overview of our approach} \label{sec:approach}

Given an {\em original code} \code{Origin}, we want to infer the 
{\em refactored code} \code{Stream} such that, for any initial program state
$S_i$, \code{Origin} and \code{Stream} produce the same final
state, i.e., they are observationally equivalent.
We consider a {\em program state} to consist of  
assignments to all the scalar variables plus a
heap representation mapping all the Java reference variables 
to their corresponding heap addresses.
%
%

Then, starting with a nondeterministic 
state $S_i$ (we use the notation $S_i{=}*$), every terminating trace according to the original code must end up in the same state
reached by applying the refactored code
(we discuss non-terminating behaviours in the last
paragraph of Sec.~\ref{sec:prog.synthesis}).
There are two things to be noted here:
(1) The initial state only considers variables that
are accessed by \code{Origin} as opposed to variables
that are live at the beginning of \code{Origin}
(we explain this in more detail in Sec.~\ref{sec:aliasing}.)
(2) We overapproximate the context of the initial code
in the sense that we may consider more initial states than
those reachable at the start of \code{Origin} in the user code.  
As a consequence, we obtain general refactorings (see Sec.~\ref{sec:preliminaries}).
%
%
%
%
Next, we explain the main steps of our refactoring procedure:

{\bf(i)} Given the original code and a nondeterministic initial state as inputs, 
we generate constraints characterising the post-state $S_f$ of the original code.
%
%
As $\code{Origin}$ contains potentially unbounded loops with external
iteration, $S_f$ is not straightforward to compute. We address this by
assuming the existence of safety invariants and generating constraints
over them.  If we consider a generic loop with a pre- and post-state
$S_i$ and $S_f$, respectively, guard $G$ and transition relation $T$:
$\{\mathit{S_i}\}$\,\code{while}($G$)\,$T$\,$\{\mathit{S_f}\}$,
%
we generate the following constraints showing that
any terminating
execution starting in a state satisfying $\mathit{S_i}$ reaches a
state satisfying $\mathit{S_f}:$
%
{\small
\begin{align}
  \exists S_f, Inv. \forall x, x' .  & \mathit{S_i(x)} \rightarrow Inv(x) ~ \wedge \label{safety_base}\\
  & Inv(x) \wedge G(x) \wedge T(x, x') \rightarrow \mathit{Inv}(x') ~ \wedge \label{safety_inductive}\\
  & Inv(x) \wedge \neg G(x) \rightarrow \mathit{S_f(x)} \label{safety_safe}
\end{align}
}
In this formula, (\ref{safety_base}) ensures that the safety invariant holds
in any state satisfying $S_i$, (\ref{safety_inductive}) checks that the invariant is
inductive with respect to the transition relation, i.e.~the transition
relation maintains the invariant, and (\ref{safety_safe}) ensures that the
invariant establishes $S_f$ on exit from the loop.  This can be
generalised to multiple, potentially nested, loops.
%
%
The two existentially quantified second-order entities, $S_f$ and $Inv$, are synthesised
in the next step.

%
{\bf(ii)} We provide the constraints generated at the previous step to our program
synthesiser (see Sec.~\ref{sec:prog.synthesis}), which outputs
$S_f$ and 
the necessary safety invariants.
Note that $S_f$ is synthesised such that it assigns all 
the scalar and reference variables and, the language
in which is synthesised only contains operations that directly
capture the semantics of the Java Stream interface.
%
%
Consequently, $S_f$ captures the semantics of the refactored code and
\code{Stream} can be generated directly from $S_f$ through a
one-to-one translation.

Essentially, our approach consists of computing the
strongest postcondition of the original code in a language capturing
the semantics of the Java Stream interface.  If we do manage to
find such a postcondition, then a refactoring exists and 
it is {\em guaranteed} to be equivalent to the original
one by construction.

\paragraph{Logical encoding} 
In order to generate the constraints at point (i), we must identify a
logical encoding for our analysis, which we use to express
$\mathit{Inv}$ and $S_f$.  
Our logic must have the ability to express:
(1)~operations supported by the Java Collection interface,
(2)~operations supported by the Java Stream interface, as well as
(3)~equality between collections (for lists this implies that we must
be able to reason about both content of lists and the order of
elements).

For this purpose, we define the Java Stream Theory (\logic). Due to lack
of space, we only provide an informal presentation of \logic in
in Fig.~\ref{fig:JCT}
containing only the operations used in the examples in the
paper.
We make use the
notion of incomplete collection/list represented by a {\em list segment}
$\hls{x}{y}$, i.e., the list starting at the node pointed by $x$ and ending
at the node pointed by $y$.

Throughout the paper we take the liberty of
referring to collections as {\em lists}. 
Also note that we capture side-effects by explicitly naming the current heap
-- heap variables $h, h'$ etc. are being introduced (as a front-end transformation), 
denoting the heap in which
each function is to be interpreted.  The mutation operators (e.g.  $get$,
$add$, $set$, $remove$) then become pure functions mapping heaps to heaps. 


\begin{figure*}[!hbt]
{\small
  \begin{tabular}{l}
    $h'=add(h, x, i, v)$:  obtain $h'$ from $h$ by inserting value $v$ at position $i$ in the list pointed\\
    by $x$.\\
$h'=add\_last(h, x, v)$:  equivalent to $add(h, x, size(h, x, \hnull), v)$\\
    $h'= set(h,x, i, v)$:  obtain $h'$ from $h$ by setting the value of the $i$-th element in the list\\
    pointed by $x$ to $v$.\\
    $h'=filter(h,x,y,\lambda v.P(v),ret)$:  obtain $h'$ from $h$ by creating a new list $ret$
    containing\\ all the elements in the list segment $\hls{x}{y}$ that match the predicate $P$.\\
$max(h,x,y)$: what is the maximum value stored in the list segment $\hls{x}{y}$?\\
    $h'=map(h,x,y,\lambda v.f(v),ret)$: obtain $h'$ from $h$ by applying the mapping function $f$ to\\
    each value in the list segment $\hls{x}{y}$ and storing the result in the list pointed by $ret$.\\
$h'=skip(h,x,y,done,n,ret)$: obtain $h'$ by creating a new list $ret$ containing the\\ remaining elements of the list segment $\hls{x}{y}$ after discarding the first $n$ elements ($done$\\ denotes the number of elements that were already skipped).\\
\end{tabular}
}
\caption{Informal Description of \logic.}
\label{fig:JCT}
\end{figure*}

\subsection{Discussion on aliasing} \label{sec:aliasing}
In the overview of our approach (Sec.~\ref{sec:approach}),
when expressing the pre-state at the beginning of \code{Origin},
we only consider the variables (and collections) that are accessed by
\code{Origin} (as opposed to all the live program variables).
Thus, one might wonder if there aren't any side-effects due to aliasing 
that we are not considering.
The answer is no, our approach is safe for reference variables as well as
the only two potential aliasing scenarios involving a 
reference variable $p$ that is not directly used by \code{Origin},
which are the following:

\begin{itemize}
\item[1.] $p$ points to a collection that is modified by \code{Origin}.  As~the
\code{Stream} refactoring is going to perform an equivalent transformation
in-place, the refactoring will be transparent to~$p$.
\item[2.] $p$ is an iterator over a collection accessed by \code{Origin}.  Then,
if \code{Origin} modifies the collection, so will \code{Stream}, which will
result in $p$ being invalidated in both scenarios.  Contrary, if \code{Origin} 
does not modify the collection, neither will \code{Stream}, and $p$ will not
be affected in either one of the cases.
\end{itemize}


Next, we illustrate scenario 1 by 
considering again method \code{removeNeg} in
Fig.~\ref{ex:filter1} with the following calling context, where we assume
$p$ points to some list and we create an alias~$p'$ of~$p$:

\begin{lstlisting}[mathescape=true,showstringspaces=false]
  ArrayList<Integer> p' = p;
  removeNeg(p);
\end{lstlisting}

At a first glance, a potential refactoring for \code{removeNeg} is: 
\begin{lstlisting}[mathescape=true,showstringspaces=false]
  l = l.stream().filter(el -> el>=0)
       .collect(toList());
\end{lstlisting}

However, this is incorrect when using the refactored function in the
calling context mentioned above: While the
list $p$ points to is correct, the list pointed by $p'$ is not updated. Thus,
after the call to \code{removeNeg}, $p$ will correctly point to the filtered
list, whereas $p'$ will continue pointing to the old unfiltered list.
To avoid such situations, we perform refactorings of code that mutates
collections in-place.  Thus, a correct refactoring for method
$\mathit{removeNeg}$ is:

\begin{lstlisting}[mathescape=true,showstringspaces=false]
  ArrayList<Integer> copy = new ArrayList<>(l);
  l.clear();
  copy.stream().filter(el -> el>=0)
               .forEachOrdered(l::add);
\end{lstlisting}

Here, we first create a copy $\mathit{copy}$ of $l$.  After performing the
filtering on $\mathit{copy}$, we use $\mathit{forEachOrdered}$, provided by
the Stream API, to add each element of the temporary stream back to the list
pointed to by $l$ (in the order encountered in the stream).  Thus, we are not
creating a new list with a new reference, but using the original one, which
makes the refactoring transparent to the rest of the program, regardless of
potential aliases.



\section{Motivating Examples} \label{sec:running-ex}
In this section, we illustrate our refactoring procedure 
on two examples. 

\paragraph{First example} We start with the one in Fig.~\ref{ex:stream},
where we create a new list by multiplying by 2 each positive value in
the list \code{list}.
As aforementioned, we must first introduce heap variables that
capture the side-effects. For this purpose, we will use the following
naming convention: the heap before executing the code (i.e., the
initial heap for both the original and the refactored code) is called
$h_i$. All the other heaps manipulated by the original program have
subscript~$o$. 

\begin{lstlisting}[mathescape=true,showstringspaces=false]
  Iterator<Integer> it = iterator(h_i,list);
  List<Integer> newList;
  h_o = new ArrayList<Integer>(h_i, newList);
  while (hasNext(h_o, it)) {
    int (el, h_o) = next(h_o, it).intValue();
    if (el > 0) {
      h_o = add_last(h_o, newlist, 2 * el);}}
\end{lstlisting}


For this example, the post-state after the execution of the original code is
captured by:
\begin{align*}
 S_f(h_i,h_o,\mathit{list},\mathit{newlist})  = \exists \mathit{list}', h_o'.  
  &h_o' {=} \mathit{filter}(h_i, \mathit{list}, \hnull, \lambda v. v>0, \mathit{list}')~\wedge\\
  &h_o {=} \mathit{map}(h_o', \mathit{list}', \hnull, \lambda v. 2 {\times} v, \mathit{newlist})
\end{align*}
%
%
The above says that the heap $h_o$ generated by the original code is
equivalent to the heap generated by applying $\mathit{filter}$ and
$\mathit{map}$ to $h_i$.  Then, the safety invariant required to prove
$S_f$ is identical with $S_f$ with the exception that it
considers that the list pointed by $\mathit{list}$ has only been partially
processed (up to the iterator $\mathit{it}$):
\begin{align*}
  \mathit{Inv}(h_i,h_o,\mathit{list},\mathit{newlist},it)  = \exists \mathit{list}', h_o'.
  & h_o' {=} \mathit{filter}(h_i, list, it, \lambda v. v>0, list') ~\wedge \\
& h_o {=} \mathit{map}(h_o', list', \hnull, \lambda v. 2 {\times} v, \mathit{newlist}))
\end{align*}

Note that, as $\mathit{filter}$ only processes the original list up to the
iterator, $\mathit{map}$ will consider the whole list generated by $\mathit{filter}$.

As $S_f$ directly captures the stream semantics,
from $S_f$ we generate 
stream code (see Fig~\ref{ex:stream} (b)).

\paragraph{Second example}
In this example, we illustrate an aggregate refactoring, as well as the
importance of checking equivalence between heap states.
For this purpose, we use the code below, where we 
compute the sum of all the elements in the list pointed-to by $l$, while 
at the same time removing from the list pointed-to by $p$ 
a number of elements equal to the size of $l$.

\begin{lstlisting}[mathescape=true]
  Iterator<Integer> it = p.iterator();  
  int sum = 0;
  for(i = 0; i<l.size(); i++) {
    sum += l.get(i);
    if(it.hasNext()) {
      it.next();
      it.remove();}}
\end{lstlisting}

%
If we were to only verify that the scalar variables after executing the
original and the refactored code, respectively, are equal, and omit checking
heap equivalence, then the following refactoring would be considered
correct:

\begin{lstlisting}[mathescape=true]
  sum = l.stream().reduce(0, (a b)->a+b);
\end{lstlisting}

This refactoring ignores the modifications performed to list $p$ and only
computes the sum of elements in the list pointed-to by $l$.  In our case, we
correctly find this refactoring to be unsound as the heap state
reached after executing the original code (where $p$ points to a modified
list) is not equivalent to the one reached after executing this refactoring
(where $p$ points to the unmodified list).  Instead, we find the following
refactoring, where we correctly capture the mutation of~$p$:

\begin{lstlisting}[mathescape=true]
  sum = l.stream().reduce(0, (a b)->a+b);
  ArrayList<Integer> copy = new ArrayList<>(p);
  p.clear();
  copy.stream().skip(l.size())
               .forEachOrdered(p::add);
\end{lstlisting}

\section{Synthesising Refactorings} \label{sec:prog.synthesis}







We compute the postcondition $S_f$ and safety invariants by using a
program synthesis engine.  Such engines are used increasingly in program
verification~\cite{DBLP:conf/cav/0001A14,DBLP:conf/lpar/DavidKL15}.  Our
program synthesiser makes use of Counter-Example Guided Inductive Synthesis
(CEGIS)~\cite{sketch} for stream refactoring.  We~present its general
architecture followed by a description of the parts specific to refactoring.

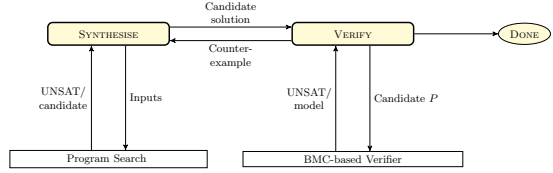
\begin{figure}
\centering
\resizebox{.6\textwidth}{!}{
 \begin{tikzpicture}[scale=0.3,->,>=stealth',shorten >=.2pt,auto,
 semithick, initial text=, ampersand replacement=\&,]

  \matrix[nodes={draw, fill=none, shape=rectangle, minimum height=.2cm, minimum width=.2cm, align=center
},
          row sep=1cm, column sep=1cm] {
   \node[box] (synth) {{\sc Synthesise}};
   \&
   \node[box] (verif) {{\sc Verify}};
   \&
   \node[ellipse, fill=yellow!20] (done) {{\sc Done}}; \\

   \& \\
   \& \\
   \node (gp) {~~~~~~~~~~~~~~Program Search~~~~~~~~~~~~~~~};
   \&
   \node (bmc) {~~~~~~~~~~~~~~~BMC-based Verifier~~~~~~~~~~~~~~~};
   
    \\
  };

   \path
    ([yshift=2em]synth.east) edge node[align=center] {Candidate\\ solution} ([yshift=2em]verif.west)
    ([yshift=-2em]verif.west) edge node[align=center] {Counter-\\example} ([yshift=-2em]synth.east)
    ([xshift=5em]verif.south) edge node {Candidate $P$} ([xshift=5em]bmc.north)
    ([xshift=-5em]bmc.north) edge node[align=center]  {UNSAT/\\model} ([xshift=-5em]verif.south)
    (verif) edge node {} (done)
    ([xshift=5em]synth.south) edge node[align=center] {Inputs} ([xshift=5em]gp.north)
    ([xshift=-5em]gp.north) edge node[align=center] {UNSAT/\\candidate} ([xshift=-5em]synth.south);
 \end{tikzpicture}
}
 \caption{The refactoring refinement loop. \label{fig:refactoring-refinement}}
\end{figure}

\paragraph{General architecture of the program synthesiser}
The design of our synthesiser is given in
Fig.~\ref{fig:refactoring-refinement} and consists of two phases, {\sc
Synthesise} and {\sc Verify}.  We will illustrate each of these phases by
using as running example the first motivational example in
Sec.~\ref{sec:running-ex}.  Our goal is to synthesise a solution
$(S_f, Inv)$. 

We start with a vacuous {\sc synthesise} phase,
where we generate a random candidate solution,
which we pass to the {\sc verify} phase. For this example,
let's assume that the random solution says that the heap manipulated by the original code
is the same 
as the initial one (i.e. the original code does not affect the heap):
$S_f(h_i,h_o,list,newlist) = h_o{=}h_i$.

In the {\sc verify} phase, we check whether the candidate solution is indeed a 
true solution for our synthesis problem (then we are ``{\sc Done}''), 
or compute a counterexample.
%
We~find such a counterexample by building a program $P_\mathit{verif}$, on
which we run Bounded Model Checking
(BMC)~\cite{DBLP:journals/ac/BiereCCSZ03}.  
BMC employs symbolic execution to map program semantics to a SAT instance
~\cite{cbmc} which verifies our equivalence constraints.
If we manage to prove partial
correctness of $P_\mathit{verif}$, then we are done.  Otherwise, we provide
the counterexample returned by BMC to the {\sc synthesise} phase.  Note that
it is sound to use BMC because the program $P_\mathit{verif}$ does not
contain loops as it uses loop invariants.  For the running example, BMC
returns a counterexample with initial heap~$h_{ce}'$ where the candidate
$S_f$ is not a true postcondition when $\mathit{list}$ contains
value~1 (added at position~0 through
$\mathit{add}(h_{ce}, \mathit{list}, 0, 1)$):
$h_{ce} = \mathit{new}(h_i, \mathit{list}) ~\wedge \enspace h_{ce}' = \mathit{add}(h_{ce}, list, 0, 1)$.



Next, in the {\sc synthesise} phase, we add the counterexample from
the previous phase to Inputs and search for a new candidate solution
by constructing a program $P_\mathit{synth}$ on which we run in parallel BMC
and a genetic algorithm (GA) to find a new candidate solution that holds for
all the Inputs.  GA simulates an evolutionary process using selection,
mutation and crossover operators.  Its fitness function is determined by the
number of passed tests.  GA maintains a large population of programs which
are paired using crossover operation, combining successful program features
into new solutions.  In~order to avoid local minima, the mutation operator
replaces instructions by random values at a comparatively low probability. 
Moreover, we use a biased crossover operation, selecting parents that solve
distinct counterexample sets for reproduction.
We use the result of either BMC or GA, depending on which one returns first. 
Again, it is sound to use BMC as the program $P_\mathit{synth}$ does not
contain loops.  For the running example, BMC returns first with a candidate
solution saying that the heap $h_o$ after the stream code is the following
(for brevity, we omit the invariant which is very similar to $S_f$):
%
$ h_o' = \mathit{filter}(h_i, \mathit{list}, \hnull, \lambda v. \text{true}, \mathit{list}')~\wedge  h_0 = \mathit{map}(h_o', \mathit{list'}, \hnull, \lambda v.2 {\times} v, \mathit{newlist})$

%
This solution is almost correct, apart from the {\em filter} predicate,
which does no actual filtering as the predicate is $true$.  Returning to the
{\sc verify} phase, we find one further counterexample denoting a list with
value~0 (which should be filtered out but it isn't):
$h_{ce} {=} \mathit{new}(h_i, list) ~\wedge \enspace h_{ce}' {=} \mathit{add}(h_{ce}, list, 0, 0)$.

Back in the {\sc synthesise} phase, this counterexample refines the $filter$ predicate,
leading to the next solution:\\
$h_o' = \mathit{filter}(h_i, \mathit{list}, \hnull, \lambda v. v \neq 0, \mathit{list}') ~\wedge
h_o = \mathit{map}(h_o', \mathit{list}, \hnull, \lambda v.2 {\times} v, \mathit{newlist})$

Still not matching the original algorithm, the {\sc verify} phase provides one
final counterexample (a list containing value $-2$ that should be filtered out, but it isn't):
$$h_{ce} {=} \mathit{new}(h_i, list) ~\wedge \enspace h_{ce}' {=} \mathit{add}(h_{ce}, list, 0, -2)$$

In the final {\sc synthesise} phase we get the solution provided in
Sec.~\ref{sec:running-ex}.

\paragraph{Elements specific to stream refactoring}
In order to use program synthesis for stream refactoring, we required 
the following:

(i) {\em The target instruction set is \logic restricted to the stream operations},
which requires 
both the {\sc verify} and {\sc synthesise} phases in the program synthesiser 
to support the \logic transformers.  
\logic directly models Java Streams such that, 
once the synthesiser finds a postcondition $S_f$, 
we only require very light processing to generate valid Java Stream code.
In particular, this processing involves the stream generation (see examples below). 

Some examples of the generated stream code are provided below, where
the LHS denotes either the post-heap $h_o$  or 
some other scalar variable $r$  captured by $S_f$ 
(expressed in \logic), and the RHS represents the corresponding stream refactoring.
For illustration, in the first example, 
after the synthesiser finds that $h_o$ in $S_f$ is 
$h_o'{=}filter(h_i,l, \hnull,\lambda v.P(v),l')$,
we generate the stream refactoring by 
adding the stream generation $l.stream()$ 
before the stream filtering $filter(\lambda v.P(v))$.

Note that $\equiv$ stands for reference equality.
This means that, as shown in Sec.~\ref{sec:aliasing}, 
we must generate Java code that modifies the original collection in place.
%

{\small
\begin{alignat*}{2}
&  h_o'{=}filter(h_i,l, \hnull,\lambda v.P(v),l') \Rightarrow l' \equiv l.stream().filter(\lambda v.P(v))\\
& h_0{=}sorted(h_i,l, \hnull,l') \Rightarrow l'\equiv l.stream().sorted()    \\
& h_0{=}skip(h_i,l,\hnull, k, 0, l')    \Rightarrow l'\equiv l.stream().skip(k) \\
& r {=} forall(h_i,l, \hnull, \lambda v.P(v))  \Rightarrow r {=} l.stream().allMatch(v \rightarrow P(v)) \\
& r {=} max(h, l, \hnull)   \Rightarrow  r {=} l.stream().max()
\end{alignat*}
}

(ii) {\em The search strategy}: we parameterise the solution language, 
where the main parameter is the length of the
solution program, denoted by $l$.  At each iteration we synthesise
programs of length exactly $l$.  We start with $l = 1$ and increment
$l$ whenever we determine that no program of length $l$ can satisfy
the specification.  When we do successfully synthesise a program, we
are \emph{guaranteed that it is of minimal length} since we have
previously established that no shorter program is correct.
%
This is particularly useful for
our setting, where we are biased towards short refactorings (see
Sec.~\ref{sec:threats.to.validity}).

\paragraph{Terminating and exceptional behaviour}
Next, we discuss how our refactoring interacts with non-terminating and exceptional 
behaviours of the original code.

%
If the original code throws an exception, then the same happens for our
modelling, and thus we fail to find a suitable refactoring.
%
The non-terminating behaviour can be due to either iterating over a collection 
with an unbounded number of elements or to a bug in the code that does not properly 
advance the iteration through the collection. 
Regarding the former, we assume that the code to be refactored handles only collections with a 
bounded number of elements. 
With respect to the second reason for non-termination, if such a bug exists in the 
original code, then it will also exist in our modelling. Thus, 
we will fail to find a suitable refactoring.

\section{Experiments}

\paragraph{Benchmark Selection}
We provide an implementation of our refactoring decision procedure, which we
have named \tool. We employed the GitHub Code Search to find relevant Java
classes that contain integer collections with refactoring opportunities to
streams.  \tool currently supports refactorings from Java external
iterators to Streams for integer collections only.  This limitation
is not conceptual, but rather due to our Java front-end based on
CBMC~\cite{cbmc}, which will be extended in future work.
The queries were specified conservatively as to not exceed the
CBMC front-end capabilities and we manually ruled out search results which
cannot be implemented using the Java~8 Stream specification.
%
We used the following search queries on 8/8/2016:
\begin{itemize}
  \item \code{List<Integer>+for+if+break++language\%3AJava&type=Code}
  \item \code{List<Integer>+while+it+remove&type=Code}
  \item \code{List<Integer>+while+add}
\end{itemize}
We found 50 code snippets with
loops from the results that fit these restrictions.

\paragraph{Experimental Setup}
In~order to validate our
hypothesis that semantics-driven refactorings are more precise than
syntax-driven ones, we compare \tool against the Integrated Development
Environments IntelliJ~IDEA~2016.1\footnote{https://www.jetbrains.com/idea/}
and NetBeans~8.2\footnote{https://netbeans.org/}, as well as against
LambdaFicator by Franklin et al.~\cite{DBLP:conf/icse/FranklinGLD04}.
These tools all provide a ``Replace with collect'' refactoring, which
matches Java code against pre-configured external iteration patterns and
transforms the code to a stream expression if they concur.  We~manually inspect
each transformation for both tools to confirm correctness.  
Since \tool's
software synthesis can be a time-consuming process, we impose a time limit
of 300\,s for each benchmark.  
All experiments were run on a 12-core
2.40\,GHz Intel Xeon E5-2440 with 96\,GB of RAM.

\paragraph{Genetic Algorithm Configuration}
We implemented a steady state genetic algorithm implementation in CEGIS, whose
fitness function is determined by the number of passed tests.
We employ a biased crossover operation, selecting parents which solve distinct
counterexamples in the CEGIS counterexample set for reproduction.  The intent is to have
parent refactorings which work for distinct intput sets produce offspring which behave correctly
for both input sets. The population size, replacement and mutation rates are configurable and
were set to 2000, 15\% and 1\% respectively for our experimental evaluation.

\paragraph{Results}\label{experiments-results}
Our results show that \tool outperforms IntelliJ, NetBeans~8.2 and LambdaFicator
by a significant margin: \tool finds 39 out of 50 (78\%) possible refactorings, whereas
IntelliJ only transforms 10 (20\%) and both NetBeans~8.2 and LambdaFicator transform 11 
benchmarks (22\%) successfully.
IntelliJ, NetBeans and LambdaFicator
combined find 15 (30\%) refactorings.  This is due to the fact
that there are many common Java paradigms, such as \code{ListIterator} or
\code{Iterator::remove}, for which none of the tools contain pre-configured
patterns and thus have no way of refactoring.  The fact that none of the pattern-based tools provide for
these situations suggests that it is impractical to try to enumerate every possible
refactoring pattern in IDEs.

If the pattern-based tools find a solution, they transform the program safely and
instantaneously, even in cases where \tool fails to synthesise a refactoring
within the allotted time limit.  Where \tool synthesised a valid refactoring, it did
so within an average of 8.5\,s.  It is worth mentioning that the syntax-driven tools
and \tool complement each other very well in our experiments, which is illustrated by the
fact that both approaches combined would have solved 44 out of the 50 refactorings (88\%)
correctly.  Loops which match the expected patterns of syntax-driven tools are handled
with ease by such tools, regardless of semantic complexity.  \tool on the other hand abstracts away
even stark syntactical differences and recognizes equivalent semantics instead, but is
limited by the computational complexity of its static analysis engine.

\tool's maximum memory usage (heap+stack) was 125MB over all benchmarks according to
valgrind massif.  We found that the majority of timeouts for \tool are due to an
incomplete instruction set in the synthesis process.  We~plan to implement missing
instructions as the program progresses out of its research prototype phase
into an industrial refactoring tool set.  A link to all benchmarks used in
the experiment is provided in the
footnote\footnote{\url{https://drive.google.com/open?id=0ByIexo3Z5N91ZlNFZTNpdU5USjQ}}.

\subsection{Threats to Validity}
\label{sec:threats.to.validity}

Our hypothesis is that we have given exemplary evidence that semantics-based
refactoring can be soundly applied, are more precise and enable more complex
refactoring schemata. As we use program analysis technology, all standard
threats to validity in this domain apply here as well; we~summarise these
only briefly.
\paragraph{Selection of benchmarks}
Our claim relates to ``usual'' programs written by human programmers, and
our results may be skewed by the choice of benchmarks.  We address this concern
by collecting our benchmarks from GitHub, which hosts a representative and
exceptionally large set of open-source software packages.  Commercial
software may have different characteristics, was not covered by our
benchmarks, and thus our claim may not extend to commercial, closed-source
software.  Furthermore, all our benchmarks are Java programs, and our claim
may not extend to any other programming language.  We focused
our experimental work on the exemplar of refactoring iteration over
collections, and our technique may not be more widely applicable.  Finally,
our Java front-end is still incomplete, only supporting lists of integers
and lacking models for many Java system classes.  This restricts our selection
to a subset of the benchmarks in our GitHub search results, which may be
biased in favour of our tool.  We~will address this issue by extending the
front-end to accept additional Java input.

\paragraph{Quality of refactorings}

Refactorings need to generate code that remains understandable and
maintainable.  Syntax-driven refactoring has good control over the resulting
code; the code generated by our semantic method arises from a complex search
procedure, and may be difficult to read or maintain.
It is difficult to assess how well our technique does with respect to this
subjective goal.  Firstly, we conjecture that small refactorings are
preferable to larger ones (measuring the number of operations).  Our method
guarantees that we find the shortest possible refactoring due to the way we
parameterise and search the space of candidate programs (as described in
Sec.~\ref{sec:prog.synthesis}). It is unclear whether human programmers
indeed prefer the shortest possible refactoring.
Secondly, our method can exclude refactorings that do not improve
readability of the program.  For instance, we exclude transformations that
include only \emph{peek} and \emph{foreach},
which are offered by the Stream~API.  A~refactoring that uses
these transformers can be trivially applied to virtually
any loop processing a collection in iteration order but is generally
undesirable.
%
%
Finally, we manually inspected the refactorings obtained with our tool and
found them to represent sensible transformations.

\paragraph{Efficiency and scalability of the program synthesiser}

We apply heavy-weight program analysis. This implies that our broader claim
is threatened by scalability limits of these techniques.  The scalability of
our particular refactoring procedure is gated by the program synthesiser. 
While for the majority of our experiments the synthesiser was able to find a
solution quickly, there were a few cases where it failed to find one at all. 
The problem was that the {\sc synthesise} portion of the CEGIS loop failed
to return with a candidate solution. Different instruction sets for the
synthesis process can help mitigate this effect.

\paragraph{Better syntax-driven refactoring} Our hypothesis relates
semantics-driven to syntax-driven refactoring.  While we have
undertaken every effort to identify and benchmark the existing syntax-driven
refactoring methods, there may be means to achieve comparable or better
results by improving syntax-driven refactoring.

\section{Related Work}

\paragraph{Program refactoring}

Cheung et al.~describe a system that automatically transforms fragments of
application logic into SQL queries~\cite{DBLP:conf/pldi/CheungSM13}. 
Moreover, similar to our approach, the authors rely on synthesis technology
to generate invariants and postconditions that validate their
transformations (a~similar approach is presented
in~\cite{DBLP:conf/cc/IuCZ10}).  The main difference (besides the actual
goal of the work, which is different from ours) to our work is that the
lists they operate on are immutable and do not support operations such as
remove.  Capturing the potential side effects caused by such operations is
one of our work's main challenges.

Syntax-driven refactoring base program transformation decisions
on observations on the program's syntax tree.  Visser
presents a purely syntax-driven framework~\cite{stratego}.  The
presented method is intended to be configurable for specific
refactoring tasks, but cannot provide guarantees about semantics
preservation.  The same holds for~\cite{txl} by Cordy et al.,
\cite{sawin} by Sawin et al., \cite{bae} by Bae et al.~and
\cite{chris} by Christopoulou et al.  In contrast to these approaches,
our procedure constructs an equivalence proof before transforming the
program. In \cite{conf/sigsoft/GyoriFDL13}, Gyori et al. present a
similar refactoring to ours but performed in a syntax-driven manner.
Steimann et al.~present Constraint-Based Refactoring in \cite{Steimann2011},
\cite{Steimann2012Pilgrim} and \cite{Steimann2011KollePilgrim}. Their approach
generates explicit constraints over the program's abstract syntax tree to
prevent compilation errors or behaviour changes by automated refactorings.
The approach is limited by the information
a program's AST provides and thus favours conservative implementations of
syntax-focused refactorings such as \emph{Pull Up Field}.
Fuhrer et al.~implement a type constraint system to introduce missing type
parameters in uses of generic classes (cf. \cite{DBLP:conf/ecoop/FuhrerTKDK05})
and to introduce generic type parameters into classes which do not provide
a generic interfaces despite being used in multiple type contexts
(cf. \cite{DBLP:conf/icse/KiezunETF07}).
%
%
%
O'Keffe and Cinn{\'{e}}ide present search-based
refactoring~\cite{search1, search2}, which is similar to syntax-driven
refactoring.  They rephrase refactoring as an optimisation problem,
using code metrics as fitness measure.  As such, the method optimises
syntactical constraints and does not take program semantics into
account.
%
%
Kataoka et al. interpret program semantics to apply refactorings
\cite{Kataoka:2001:ASP:846228.848644}, but use dynamic test execution
rather than formal verification, and hence their transformation lacks
soundness guarantees.
Franklin et al. implement a pattern-based refactoring approach
transforming statements to stream queries~\cite{Gyori:2013:CGI:2491411.2491461}.
Their tool LambdaFicator~\cite{DBLP:conf/icse/FranklinGLD04} is available as a
NetBeans branch. We compared \tool against it in our experimental evaluation
in Sec.~\ref{experiments-results}.
%
%

\paragraph{Program synthesis}

An approach to program synthesis very similar to ours is Syntax Guided
Synthesis (SyGuS)~\cite{sygus}.  SyGuS synthesisers supplement the logical
specification with a syntactic template that constrains the space of allowed
implementations.  Thus, each semantic specification is accompanied by a
syntactic specification in the form of a grammar.  Other second-order
solvers are introduced in~\cite{DBLP:conf/pldi/GrebenshchikovLPR12,
DBLP:conf/cav/BeyenePR13}.  As opposed to ours, these focus on
Horn clauses.

\section{Conclusion}

We conjecture that refactorings driven by the semantics of programs have
broader applicability and are able to address more complex refactoring
schemata in comparison to conventional syntax-driven refactorings, thereby
increasing the benefits of automated refactoring.  The space of possible
semantic refactoring methods is enormous; as an instance, we have presented
a method for refactoring iteration over Java collection classes based on
program synthesis methods.  Our experiments indicate that refactoring using
this specific instance is feasible, sound and sufficiently performant. 
Future research must broaden the evidence for our general hypothesis by
considering other programming languages, further, ideally more complex
refactoring schemata, and other semantics-based analysis techniques.



\bibliographystyle{abbrv}
\bibliography{document}{}




\appendix

\section{Another motivational example}
Next, we provide a more involved example where the original code has nested
loops.  For this purpose we use the code for selection sort in
Fig.~\ref{ex:sort} (a).  First, we introduce the heap variable as shown in
Fig.~\ref{ex:sort} (b).
Also, we provide in Fig.~\ref{fig-appendix:JCT} more of the \logic operations
(especially those used by the example that were not provided in Fig.~\ref{fig:JCT}).
If $\mathit{Inv}_\mathit{out}$ and
$\mathit{Inv}_\mathit{in}$ are the safety invariants for the outer and inner
loops, respectively, then the constraints for the outer loop are
(we omit the inner loop as it follows directly from the equations
(\ref{safety_base}), (\ref{safety_inductive}), (\ref{safety_safe}) in
Sec.~\ref{sec:preliminaries}):

{\small 
\begin{align}
& \forall h_i,h_o,l,j. \exists min, h_o', temp. Inv_{out}(h_i,h_o,l, 0) ~\wedge \label{vc:1}\\
& (Inv_{out}(h_i,h_o,l, j) \wedge j {<} (size(h_o,l){-}1) ~\wedge \label{vc:2}\\
& \quad Inv_{in}(h_i,h_o,l, size(h_o,l), j, min) ~\wedge \label{vc:3}\\ 
& ~~ temp {=} get(h_o,l, j) \wedge h_o' {=} set(h_o,l, j,get(h_o,l, min)) ~\wedge \label{vc:4}\\
& ~~  h_o{=} set(h_o',l, min, temp)) \Rightarrow Inv_{out}(h_i,h_o,l,j{+}1) ~\wedge \label{vc:5}\\
& ~~ Inv_{out}(h_i,h_o,l,j) \wedge j {\geq} (size(h_o,l){-}1) \Rightarrow S_f(h_i,h_o,l) \label{vc:6}
\end{align}
}

Constraint (\ref{vc:1}) says that the outer loop's invariant
must hold in the initial state, constraints (\ref{vc:2}), (\ref{vc:3}), (\ref{vc:4})
and~(\ref{vc:5}) check that $\mathit{Inv}_\mathit{out}$ is re-established by
the outer loop's body (by making use of $\mathit{Inv}_\mathit{in}$), whereas
(\ref{vc:6}) asserts that the $S_f$ postcondition must hold on exit
from the outer loop.  For this example, we find the following solution:
{\small \begin{align*} &
Inv_{out}(h_i,h_o,l, j) = \exists h_o', it_{j}.
h_o = sorted(h_o',l,it_{j},l') ~\wedge  h_o' =
getIterator(h_o,l,j,it_{j}) ~\wedge \\
& \quad max(h_o',l',it_{j})
{\leq} min(h_o',it_{j},\hnull) \\
%
& Inv_{in}(h_i,h_o,l,i,j,min) = \exists h_o',h_o''.(\hmin(h_o'',it_{j},it_{i}) {=} \code{min} ~\wedge\\
& h_o'{=}getIterator(h_o,l,j,it_{j}) \wedge h_o''{=}getIterator(h_o,l,i,it_{i}))\\
%
&S_f(h_i,h_o,l) = h_o{=}sorted(h_i,l, \hnull,l)
\end{align*}
}
The invariant of the outer loop expresses the fact that the lists
$l$ in the original code is 
sorted until element $j$.  Because~our theory \logic supports iterator-based
$sorted$ predicate (rather than index-based), we need to create
iterator $it_{j}$ to the $j$-th element in the list~$l$.
Additionally, the invariant of the outer loop
captures the fact that the maximum element in the already sorted portion of
list $l$ is at most equal to the minimum element from the portion still to
be sorted.

The inner loop's invariant captures the fact that the minimum element in the
list segment between the $j$-th and the $i$-th element is \code{min}
(program variable).  The postcondition $s_f$ captures the fact that
list $l$ is sorted in $h_o$.

\begin{figure*}
\begin{framed}
\begin{minipage}{.5\textwidth}
  \begin{lstlisting}[mathescape=true,showstringspaces=false,escapechar={§}]
void sorting(List<Integer> l) {
  int min, temp;
  for (int j = 0; j < l.size()-1; j++) {
    min = j; 
    for (int i = j+1; i < l.size(); i++)
      if (l.get(i)<l.get(min)) min = i; 
    temp = l.get(j);
    l.set(j, l.get(min));
    l.set(min, temp);}}
§\hspace{3cm}\small{(a)}§
\end{lstlisting}
\end{minipage}\\
\begin{minipage}{.45\textwidth}
  \begin{lstlisting}[mathescape=true,showstringspaces=false,escapechar={§}]
void sorting(List<Integer> l) {
  int min, temp;
  h_o = copyHeap(h_i);
  for (int j = 0; j < size(h_o,l)-1; j++) {
    min = j; 
    for (int i = j+1; i < size(h_o,l); i++)
      if (get(h_i,l,i)<l.get(h_o,l,min)) min = i;
    temp = get(h_o,l,j);
    h_o' = set(h_o, l, j, get(h_o, l, min));
    h_o = set(h_o', l, min, temp);}}
§\hspace{3cm}\small{(b)}§
\end{lstlisting}
\end{minipage}
\end{framed}
\caption{Selection sort: (a) original code (b) with explicit heap variables.}
\label{ex:sort}
\end{figure*}

From postcondition $S_f$ we generate the following refactored code, where we modify 
$l$ in-place by using a local copy. 

\begin{lstlisting}[mathescape=true,showstringspaces=false]
List<Integer> sorting(List<Integer> l){
  List<Integer> copy = new List<>(l);
  l.clear();
  copy.stream().sorted()
               .forEachOrdered(l::add);}
\end{lstlisting}

\section{Java Stream Theory} \label{sec:JCT}

\begin{figure*}[!hbt]
{\small
  \begin{tabular}{l}
 $\halias(h,x,y)$:  do $x$ and $y$ point to the same node in heap $h$? \\
$size(h,x,y)$:  what is the length of the list segment from $x$ to $y$ in $h$?\\
    $get(h,x,i)$:  what is the value stored in the $i$-th node of the list pointed by $x$ in heap $h$?\\
 $h' = \hremove(h, x)$:  obtain $h'$ from $h$ by removing the node pointed by $x$. In $h'$, $x$
and\\ all its aliases will point to the successor of the removed node.\\
$h' = removeVal(h, x,y, v)$:  obtain $h'$ from $h$ by removing the node with value $v$ from\\ the list segment $\hls{x}{y}$.\\
 $exists(h,x,y,\lambda v. P(v))$:  is there any value $v$ in the list segment $\hls{x}{y}$ such that $P(v)$\\ holds?\\
 $forall(h,x,y,\lambda v. P(v))$:  is it the case that for all values $v_1 \ldots v_n$ in the list segment\\ $\hls{x}{y}$, $P(v_1) \ldots P(v_n)$ hold?\\
$h'=sorted(h,x,y,ret)$:  obtain $h'$ from $h$ by sorting the elements stored in the list\\ segment $\hls{x}{y}$ in the list $ret$ 
($h'$ will contain both the list segment $\hls{x}{y}$ and the list $ret$).\\
$min(h,x,y)$: what is the minimum value stored in the list segment $\hls{x}{y}$?\\
$h'=limit(h,x,y,done,n,ret)$: obtain $h'$ by creating a new list $ret$ containing the\\ elements of the list segment $\hls{x}{y}$, after its length was truncated to $n$ ($done$ denotes\\ the number of elements that were dropped).\\
$reduce(h,x,y,v,\lambda a ~b. f(a,b))$: 
performs a reduction on the elements of the list\\ segment $\hls{x}{y}$, using the identity value $v$ and the accumulation function $f$, and returns\\ the reduced value.\\
$h'=concat(h,x,y,a,b,ret)$: obtain $h'$ from $h$ by creating a new list $ret$ containing\\ all the elements in the list segment 
$\hls{x}{y}$  followed by all the elements in the list segment\\ $\hls{a}{b}$.\\
$h' = copy(h,x,y,ret)$: obtain $h'$ by creating a new list $ret$ that contains the elements of\\ the list segment $\hls{x}{y}$.\\
$h' = \hnew(h, x)$ obtain $h'$ from $h$ by assigning $x$ to point to $\hnull$.\\
$equalLists(h,x,y,h',a,b)$ is list segment $\hls{x}{y}$ in heap $h$ equal to list segment  $\hls{a}{b}$ in\\ heap $h'$ (i.e., do they contain the same elements in the same order)?\\
$h' = getIterator(h,x,i,it)$ obtain heap $h'$ by creating a new iterator $it$ that points to\\ the $i$-th element in the list pointed-to by $x$.   
\end{tabular}
}
\caption{Informal Description of \logic (continuation from Fig.~\ref{fig:JCT}).}
\label{fig-appendix:JCT}
\end{figure*}

We designed \logic such that it meets several criteria:\\ 
1.~Express operations allowed by the Java Collection interface,
operations allowed by the Java Stream interface as well as equality
between collections (for lists this implies that we must be able to
reason about both content of lists and the order of elements).\\
2.~\logic must be able to reason about the content and size of
partially constructed lists (i.e., list segments), which are required
when expressing safety invariants. 
For illustration, in Fig.~\ref{ex:stream}, the safety invariant captures the fact that 
$h_s$ is obtained from $h_i$ by filtering the list segment $\hls{list}{it}$.\\
%
3.~\logic must enable concise $S_f$ postconditions and invariants as
we use program synthesis to infer these.  Thus, the smaller they are, the
easier to synthesise.\\
%
 
To the best of our knowledge, there is no existing logic that meets all the
criteria above.  The majority of recently developed decidable heap logics
are not expressive enough (fail points~1
and~2)~\cite{DBLP:conf/cav/ItzhakyBINS13, DBLP:conf/cav/PiskacWZ13,
DBLP:conf/esop/BrainDKS14, DBLP:conf/popl/MadhusudanPQ11,
DBLP:conf/atva/BouajjaniDES12, DBLP:conf/lpar/DavidKL15}, whereas very
expressive logics such as FOL with transitive closure are not concise and
easily translatable to stream code (fail point~3).

While our theory is undecidable, we found it works well for our 
particular use case.


\paragraph{Semantics.}

We first define the model used to interpret \logic formulae.  The set of
reference variables is denoted by~$PV$.  Note that, as already mentioned in
the paper, these reference variables are those accessed in the code to be
refactored (as opposed to all the reference variables in the program).

\begin{definition}[Heap] \label{def:heap}
A heap over reference variables $PV$ is a tuple $H = \langle G, L_P, L_D
\rangle$.  $G$ is a graph with vertices $V(G)$ and edges $E(G)$, $L_P : PV
\to V(G)$ is a labelling function mapping each reference variable to a vertex
of $G$ and $L_D : V(G) \to D$ is a labelling function associating each
vertex to its data value (where $D$ is the domain of the data values).
\end{definition}

Given that we are interested in heaps managed by Java Collections, we
restrict the class of models to those where each vertex has outdegree 0 or 1
(i.e.~we cannot have multiple edges coming out of a node).
We assume that the reference variables include a special name {\bf null}.

\ifdefined\extended  

Function $val(h,x)$ returns the value stored in the node pointed by $x$,
$next(h,x)$~returns a reference to the next node after the one pointed by $x$ and
it is defined as the unique vertex such that $(x, next(x)) \in E(h)$, and 
$add0(h,e,x)$ returns the heap obtained by appending element $e$ at the 
beginning of the list pointed by~$x$. For the 
latter we provide the pointwise definition:
\begin{flalign*}
& add0_V(h, e, x) \defeq V(h) \cup \{ q \} \text{\quad where $q$ is a fresh vertex} \\
& add0_{L_D}(h,e, x) \defeq L_D(h)[q \mapsto e] \\
& add0_E(h,e, x)  \defeq E(h) \cup \{ (q, L_P(h)(x)) \} 
\end{flalign*}

The semantics of \logic is defined recursively in Fig.~\ref{fig:JCT-formal}.
Note that functions $minimum$ and $maximum$ return the minimum and maximum between 
the values receives as arguments, respectively.  
While in Fig.~\ref{fig:JCT-formal} we provide the semantics for index-based operations (e.g. 
$set(x,y,i,v)$), we also support iterator-based ones (e.g. $h' {=} set(h,it,v)$ 
returns the heap obtained by setting the value of the node pointed by $it$ to $v$ in heap $h$).

\setlength{\jot}{1em}
\begin{figure*}
\begin{framed}
\begin{gather}
val(h,x) = L_D(h)(x)\\
\halias(h,x,y) \Leftrightarrow L_P(h)(x){=}L_P(h)(y)\\
\frac{i=0}{get(h,x,i) {=} val(h,x)}\\
\frac{i>0}{get(h,x,i) {=}  get(h,next(h,x),i{-}1)}\\
\frac{\halias(h,x, y)}{size(h,x,y) = 0}\\
\frac{\neg \halias(h,x, y)}{size(h,x,y) =  1 {+} size(h,next(h,x),y)}\\
\frac{\halias(h,x, y)}{\hmax(h,x,y) = -\infty}\\ 
\frac{\halias(h,x, y)}{\hmin(h,x,y) = \infty}\\
\frac{\halias(h,x, y)}{\hexists(h,x,y,\lambda v. P(v)) = false}\\
\frac{\halias(h,x, y)}{\hforall(h,x,y,\lambda v. P(v)) = true}\\
\frac{h'=copy(h,x,\hnull,l)}{add(h,x,0,v) {=} add0(h',v,l)}\\
\frac{h'=copy(h,next(h,x),\hnull,l)}{set(h,x,0,v) {=} add0(h',v,l)}\\
\frac{halias(h_1,x, y) \wedge \halias(h_2,a,b)}{equalLists(h_1,x,y,h_2,a,b) =
true}\\
\frac{\halias(h_1,x, y) \wedge \neg
\halias(h_2,a,b)}{equalLists(h_1,x,y,h_2,a,b) = false}\\
\frac{\neg halias(h_1,x, y) \wedge
\halias(h_2,a,b)}{equalLists(h_1,x,y,h_2,a,b) = false}
\end{gather}
\end{framed}
\caption{Inference rules for Java Collection Theory.}
\label{fig:JCT-formal}
\end{figure*}
\begin{figure*}\ContinuedFloat
\begin{framed}
\begin{gather}
\frac{alias(h,x,y)\wedge h'{=}h[L_p(h') {=} L_P(h) \cup \{ret\mapsto
\hnull\}]}{copy(h,x,y,ret) {=} h'}\\
\frac{\neg alias(h,x,y) \wedge h'=copy(h,next(h,x),y,ret)}{copy(h,x,y,ret) {=}
add0(h',val(h,x),ret)}\\
\frac{i>0\wedge h'=add(h,next(h,x),i{-}1,v)}{add(h,x,i,v) {=} 
add0(h',val(h,x),next(h',x))}\\
\frac{i>0 \wedge h'=set(h,next(h,x),i{-}1,v)}{set(h,x,i,v) {=}
add0(h',val(h,x),next(h',x))}\\
\frac{\halias(h,x, y)\wedge h'{=}h[L_p(h') {=} L_P(h) {\cup} \{ret{\mapsto}
\hnull\}]}{\hmap(h,x,y,\lambda v. f(v),ret) {=} h'}\\
\frac{\neg \halias(h,x, y) \wedge h'{=}\hmap(h,next(h,x),y,\lambda v.
f(v),ret)}{\hmap(h,x,y,\lambda v. f(v),ret) =  \hadd0(h', f(val(h,x)),ret)}\\
\frac{\halias(h,x,y){\wedge} h'{=}h[L_p(h') {=} L_P(h) {\cup} \{ret{\mapsto}
\hnull\}]}{skip(h,x,y,done,n,ret){=}h'}\\
\frac{\neg\halias(h,x,y) \wedge
n{>}0}{skip(h,x,y,done,n,ret){=}skip(h,next(h,x),y,done{+}1,n{-}1,ret)}\\
\frac{\neg \halias(h,x,y) \wedge n{=}0 \wedge h'{=}h[L_p(h') {=} L_P(h) {\cup}
\{ret{\mapsto} L_P(h)(x)\}]}{skip(h,x,y,done,n,ret){=}h'}\\
\frac{\halias(h,x, y)\wedge h'{=}h[L_p(h') {=} L_P(h) {\cup} \{ret{\mapsto}
\hnull\}]}{\hfilter(h,x,y,\lambda v. P(v),ret) {=} h'}\\
\frac{\neg \halias(h,x, y) \wedge \neg P(val(h,x))}{\hfilter(h,x,y,\lambda v.
P(v),ret) {=}  \hfilter(h,next(h,x),y,\lambda v. P(v),ret)}\\
\frac{\neg \halias(h,x, y) \wedge P(val(h,x))\wedge
h'{=}\hfilter(h,next(h,x),y,\lambda v. P(v),ret)}{\hfilter(h,x,y,\lambda v.
P(v),ret) =  \hadd0(h', val(h,x),ret)}\\
\frac{(\halias(h,x,y) \vee n{=}0)\wedge h'{=}h[L_p(h') {=} L_P(h) {\cup}
\{ret{\mapsto} \hnull\}]}{limit(h,x,y,done,n,ret){=}h'}\\
\frac{\neg\halias(h,x,y) \wedge n{>}0 \wedge
h'{=}limit(h,next(h,x),y,done{+}1,n{-}1,ret)}{limit(h,x,y,done,n,ret){=}add0(h',val(h,x),ret)}
\end{gather}
\end{framed}
\caption[]{Inference rules for Java Collection Theory.}
\label{fig:JCT-formal}
\end{figure*}
\begin{figure*}\ContinuedFloat
\begin{framed}
\begin{gather}
\frac{\halias(h,x, y)\wedge h'{=}h[L_p(h') {=} L_P(h) {\cup} \{ret{\mapsto}
\hnull\}]}{sorted(h,x,y,ret) = h'}\\
\frac{\neg \halias(h',x, y) \wedge
h'{=}sorted(removeVal(h,x,y,min(h,x,y)),x,y,ret)}{sorted(h,x,y,ret) = 
add0(h',min(h,x,y),ret)}\\
\frac{\neg \halias(h,x, y)}{\hmax(h,x,y) =  maximum(P(val(h,x)),
\hmax(h,next(h,x),y))}\\
\frac{\neg \halias(h,x, y)}{\hmin(h,x,y) =  minimum(P(val(h,x)),
\hmin(h,next(h,x),y))}\\
\frac{\neg \halias(h,x, y)}{\hexists(h,x,y,\lambda v. P(v)) \Leftrightarrow 
P(val(h,x)) \vee \hexists(h,next(h,x),y,\lambda v. P(v))}\\
\frac{\neg \halias(h,x, y)}{\hforall(h,x,y,\lambda v. P(v)) \Leftrightarrow  P(val(h,x))
\wedge \hforall(h,next(h,x),y,\lambda v. P(v))}\\
\frac{\halias(h,x, y)}{reduce(h,x,y,v,\lambda a ~b. f(a,b)) = v}\\
\frac{\neg \halias(h,x, y)}{reduce(h,x,y,v,\lambda a ~b. f(a,b)) =  f(v,
reduce(h,next(h,x),y,v,\lambda a ~b. f(a,b)))}\\
\frac{\neg \halias(h_1,x, y) \wedge \neg alias(h_2,a,b)}{val(h_1,x) ==
val(h_2,a) \wedge equalLists(h_1,next(h_1,x), y, h_2,next(h_2,a), b)}
\end{gather}
\end{framed}
\caption[]{Inference rules for Java Collection Theory.}
\label{fig:JCT-formal}
\end{figure*}

\else

We define the semantics of \logic recursively.  Due to lack of space, we
don't put the resulting inference rules in the paper and make them
available in the anonymised extended version \cite{extversion}.
Besides index-based operations (e.g.~$set(x,y,i,v)$), we also support
iterator-based ones (e.g.~$h' {=} set(h,it,v)$ returns the heap obtained by
setting the value of the node pointed by $it$ to $v$ in heap $h$).

\fi

One important check that we must be able to perform in order to
prove equivalence between program states is
is that of heap equivalence.
In order to define this notion 
we first assign $PV_{\cap}$ to be the set of reference variables that are used by
both the original code and the refactored one (this excludes local variables
such as iterators that are used by only one of the codes).  Then:

\begin{definition} \label{def:equiv-heaps}
Heap $h$ and $h'$ are equivalent, written as $h{=}h'$, 
iff the underlying graphs reachable from $PV_{\cap}$ are isomorphic.
\end{definition}

Intuitively, this means that all the lists in $h$ and $h'$ pointed-to by the
same variable from $PV_{\cap}$, respectively, are equal.

\paragraph{Set collections}

For our refactoring procedure, we use lists as the internal representation
for collections denoting sets, meaning that we impose an order on the
elements of sets.  While we may miss some refactorings, this procedure is
sound: if two collections are equal with respect to some ordering, they are
also equal when no order is imposed.

\end{document}

%% file: macros.tex
\newcommand{\doi}[1]{}{}

\newcommand{\defeq}{\ensuremath{\stackrel{\mathrm{def}}{=}}}

\newcommand{\hmax}{\ensuremath{\mathit{max}}\xspace}
\newcommand{\hmin}{\ensuremath{\mathit{min}}\xspace}

\newcommand{\hmap}{\ensuremath{\mathit{map}}\xspace}

\newcommand{\halias}{\ensuremath{\mathit{alias}}\xspace}

\newcommand{\hfilter}{\ensuremath{\mathit{filter}}\xspace}
\newcommand{\hnew}{\ensuremath{\mathit{new}}\xspace}

\newcommand{\hremove}{\ensuremath{\mathit{remove}}\xspace}
\newcommand{\hadd}{\ensuremath{\mathit{add}}\xspace}

\newcommand{\hexists}{\ensuremath{\mathit{exists}}\xspace}
\newcommand{\hforall}{\ensuremath{\mathit{forall}}\xspace}

\newcommand{\hls}[2]{#1 {\rightarrow^*} #2}
\newcommand{\hnull}{\ensuremath{\mathbf{null}}}

\newcommand{\code}[1]{\lstinline|#1|} 
\newcommand*{\codenospace}[1]{\lstinline|#1|\kern-1ex}%

\newcommand{\logic}{JST\xspace}
\newcommand{\tool}{Kayak\xspace}

\tikzset{
    box/.style={
           rectangle,
           rounded corners,
           draw=black, very thick,
           fill=yellow!20,
           text width=10em,
           minimum height=2em,
           text centered},
  futurebox/.style={
           rectangle,
           rounded corners,
           draw=black, very thick,
           fill=blue!10,
           text width=10em,
           minimum height=2em,
           text centered},
  longfuturebox/.style={
           rectangle,
           rounded corners,
           draw=black, very thick,
           fill=purple!20,
           text width=10em,
           minimum height=2em,
           text centered},
    arrow/.style={
           ->,
           thick,
           shorten <=2pt,
           shorten >=2pt,}
}